\pdfoutput=1

\documentclass[aps,prd,twocolumn,groupedaddress,preprintnumbers,superscriptaddress,nofootinbib]{revtex4}
\usepackage{amsmath,amssymb,latexsym}
\usepackage{graphicx}

\newcommand{\beq}{\begin{equation}}   
\newcommand{\eeq}{\end{equation}}
\newcommand{\bea}{\begin{eqnarray}}   
\newcommand{\eea}{\end{eqnarray}}
\newcommand{\bear}{\begin{array}}  
\newcommand {\eear}{\end{array}}
\newcommand{\bef}{\begin{figure}}  
\newcommand {\eef}{\end{figure}}
\newcommand{\bec}{\begin{center}}  
\newcommand {\eec}{\end{center}}
\newcommand{\non}{\nonumber}

\begin{document}

\title{
Running Spectral Index from Inflation with Modulations
}


\author{
Takeshi Kobayashi
}\email[]{takeshi.kobayashi@ipmu.jp}
\affiliation{
Institute for Cosmic Ray Research, The University of Tokyo, 5-1-5
Kashiwanoha, Kashiwa, Chiba 277-8582, Japan 
}
\author{
Fuminobu Takahashi
}\email[]{fuminobu.takahashi@ipmu.jp}
\affiliation{
Institute for the Physics and Mathematics of the Universe, The
University of Tokyo, 5-1-5 Kashiwanoha, Kashiwa, Chiba 277-8583, Japan
}

\preprint{ICRR-Report-576}
\preprint{IPMU10-0203}


\begin{abstract}
We argue that a large negative running spectral index, if confirmed,
 might suggest that there are abundant structures in the inflaton
 potential, which result in a fairly large (both positive and negative)
 running of the spectral index at all scales. It is shown that the center value
 of the running spectral index suggested by the recent CMB data can be
 easily explained by an inflaton potential with superimposed periodic
 oscillations. In contrast to cases with constant running, the
 perturbation spectrum is enhanced at small scales, due to the repeated
 modulations. We mention that such features at small scales may be seen
 by  $21$ cm observations in the future.  
\end{abstract}

\maketitle

\section{Introduction}
\label{sec:intro}

Cosmic inflation~\cite{Starobinsky:1980te,Sato:1980yn,Guth:1980zm} sets
the initial conditions for the subsequent Hot Big Bang cosmology, and
also generates the primordial curvature perturbation of the
universe. The nature of this primordial density perturbation is studied through
various cosmological observations, and especially CMB data can be well fitted by 
a power-law primordial spectrum,  which is a typical prediction of the standard slow-roll inflation.

Interestingly, however, when relaxing the assumption of a power-law spectrum, latest CMB observations
show a slight preference for a  fairly large negative running spectral index. 
Ignoring tensor modes, the WMAP 7-yr data combined with ACT 2008 data
give the bounds 
$n_s = 1.032 \pm 0.039$ and
$d n_s / d\ln k = -0.034 \pm 0.018~(68\%{\rm \, CL})$ at the pivot scale
$k_0 = 0.002\,{\rm Mpc}^{-1}$~\cite{Dunkley:2010ge}. 
Even with the tensor modes, similar bounds are obtained using the WMAP data only~\cite{Larson:2010gs}.
In the usual slow-roll inflation model with a featureless potential,  the running spectral index 
is second order in the slow-roll parameters, and so it is expected to be of $O(10^{-3})$.
Thus, although the evidence for the running spectral index
is not conclusive, it is worth studying its implications for the inflation models.

The center value of the running spectral index, if taken at face value,
suggests either the inflaton potential with a step-like feature (see
e.g. \cite{Starobinsky:1992ts,Chung:2003iu,Cline:2006db,Espinosa:2006pb,Joy:2007na,Joy:2008qd}
and references therein), or multiple
inflation~\cite{Kawasaki:2003zv,Yamaguchi:2003fp},  since otherwise
inflation would terminate within about $30$
e-foldings~\cite{Easther:2006tv}. In both cases, fine-tuning is needed
to ensure that the CMB scale exits the horizon at a specific time when
the inflaton comes close to the feature in its potential, or at about
$10$ e-foldings before the end of the first inflation.
Such fine-tuning seems a puzzle and calls for explanation.

We argue in this paper that, if there is indeed large negative running spectral index, 
this might suggest that the inflaton potential has structures
everywhere in the potential 
and such a potential may be common in the landscape of inflation models.
Such repeated modulations to the inflaton potential can account for a perturbation spectrum which is
blue tilted with negative running at the CMB scale, if one period of
the modulations is large enough to encompass the observed scales. 
Inflaton potentials with modulations can lead to repeated
features in the perturbation spectrum, in contrast to the case of
single-stepped potentials. 
See also~\cite{Feng:2003mk} for related work in the context of Natural
Inflation models \cite{Freese:1990rb,Adams:1992bn} with extra dimensions~\cite{ArkaniHamed:2003wu,ArkaniHamed:2003mz}.

Modulations to the inflaton potential can arise due to the 
microphysics governing inflation. Especially for large-field models, one
expects that the super-Planckian field ranges may contain many features
whose scale is typically related to that of the underlying 
microphysics. Explicit constructions of large-field models in string
theory~\cite{Silverstein:2008sg,McAllister:2008hb,Flauger:2009ab}
actually show that periodic modulations of the potential can show  
up, whose period is determined by e.g., the size of the internal
compactified space. (See also \cite{Tye:2008ef,Tye:2009ff} for
discussions on small-scale features along the inflaton field range in
the cosmic landscape.)

%
%
%
%

To be explicit, we consider large-field inflation from power-law
potentials with superimposed periodic modulations, and show that the center value of the running 
spectral index can be realized. What differs from the previous studies using single-stepped potentials
is that the running spectral index is not localized, but it is negative at the entire CMB scales 
and becomes positive at small scales. Such characteristic behavior of the density perturbation
may be seen by the future $21$ cm observations~\cite{Lewis:2007kz,Tegmark:2008au}. 
The potential detection of such oscillatory feature in the power spectrum by future observations was studied in
Ref.~\cite{Hamann:2008yx}, with the main focus on oscillations with (much) shorter period.


The rest of the paper is organized as follows. In Sec.~\ref{sec:basic}
we briefly explain our idea of generating the running spectral
index. Then we estimate the power spectrum of the density perturbation
based on inflation models with a linear, and then a generalized polynomial
potential superimposed with periodic modulations, both analytically and
numerically in Sec.~\ref{sec:lin} and Sec.~\ref{sec:power-law}. The last
section is devoted to discussions and conclusions.

\section{Basic idea}
\label{sec:basic}
Let us first give our basic idea in this section.
In this paper we use the units $c = \hbar = M_p =1$. 

The fact that the CMB data can be well fitted by a Gaussian power-law primordial spectrum indicates that
the standard slow-roll inflation paradigm is a valid description of the evolution of the early universe. 
However, if one looks at the details of the inflaton potential, there might be non-trivial structures which do
not change the overall behavior of the inflaton, but nonetheless affect
the tilt and/or the running spectral index 
in a non-negligible way. 

\begin{figure}[t!]
 \includegraphics[width=0.7\linewidth]{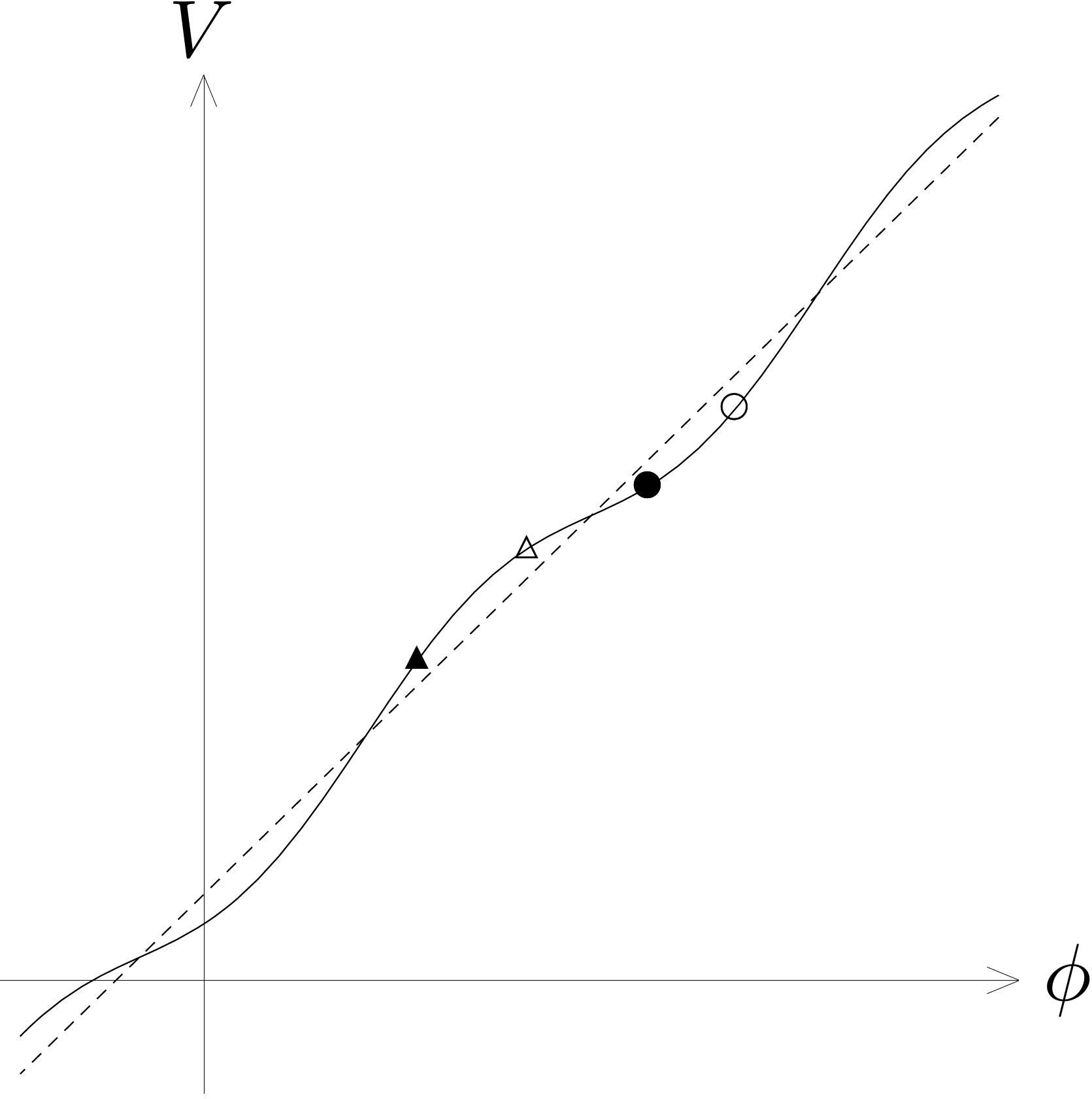}
 \caption{A sketch of the inflaton potential with superimposed periodic
 oscillations. The dashed line is a smooth potential without
 modulations. The modulations are amplified for visualization purposes. 
  The resulting curvature perturbation and its tilt at
 scales exiting the horizon as the inflaton approaches the markers
 are shown in Fig.~\ref{fig:im_sp} and Fig.~\ref{fig:im_ti}. The
 observed CMB scales are considered to lie within a half oscillation period.}
 \label{fig:im_pot}
\end{figure}

If the running of the spectral index is large at the CMB scales, 
it may suggest that the inflaton potential has step-like structures
everywhere, since severe fine-tuning would be  
needed otherwise. So we consider an inflaton potential $V(\phi)$ with
modulations such as shown in Fig.~\ref{fig:im_pot}. 
We decompose the potential as 
\beq
V(\phi) = V_0(\phi) + V_{mod}(\phi),
\eeq
where the first term is a smooth potential sourcing a (nearly) constant
spectral tilt, and the second term represents the modulations. 
We assume that the overall inflaton dynamics (such as the Hubble
parameter and the inflaton velocity) is not altered drastically by the
modulations, i.e.,
\bea
\label{con1}
|V_0(\phi)|&\gg& |V_{mod}(\phi)|,\\
\label{con2}
|V_0'(\phi)| &>& |V_{mod}'(\phi)|.
\eea
We further suppose that the slow-roll approximations 
\begin{equation}
  3 H \dot{\phi} \simeq - V', \qquad 
  3 H^2 \simeq V, 
 \label{slow-roll}
\end{equation}
are valid. 
Then, in order for the running of the spectral index to be large enough
so that the spectral tilt switches between blue and red within the
observed scales, we require
\bea
\label{con3}
|V_0''(\phi)| &\lesssim& |V_{mod}''(\phi)|,\\
\label{con4}
|V_0'''(\phi)| &\ll & |V_{mod}'''(\phi)|.
\eea
Thus, we are invoking a hierarchy
\begin{equation}
 \left|\frac{V_{mod}}{V_0}\right|  \ll
  \left|\frac{V_{mod}'}{V_0'}\right| 
 \ll \left|\frac{V_{mod}''}{V_0''}\right| \ll
 \left|\frac{V_{mod}'''}{V_0'''}\right|. \label{hier}
\end{equation}
We note that the conditions (\ref{con3}), (\ref{con4}), and (\ref{hier})
need not hold all along the inflaton potential, but are only required to
be satisfied occasionally during inflation for large running of the
spectral index to be produced.

An important assumption here is that the effect of the modulations $V_{mod}$ (and its derivatives) on the inflaton dynamics
is sufficiently small when averaged over a sufficiently long time or large field space. One example
for such modulations is a sine function, and we will see in the
following sections that the above conditions can be easily satisfied in this case. 
We emphasize here that  the effect of modulations can be missed in such an analysis that  the inflaton potential is expanded 
in terms of a power of $\phi$ up to a finite order~\cite{Easther:2006tv}. On the other hand, as we shall see, 
the spectral index and its running are easily modified by the modulations,
while not affecting the overall inflaton dynamics.

\begin{figure}[t!]
 \includegraphics[width=0.7\linewidth]{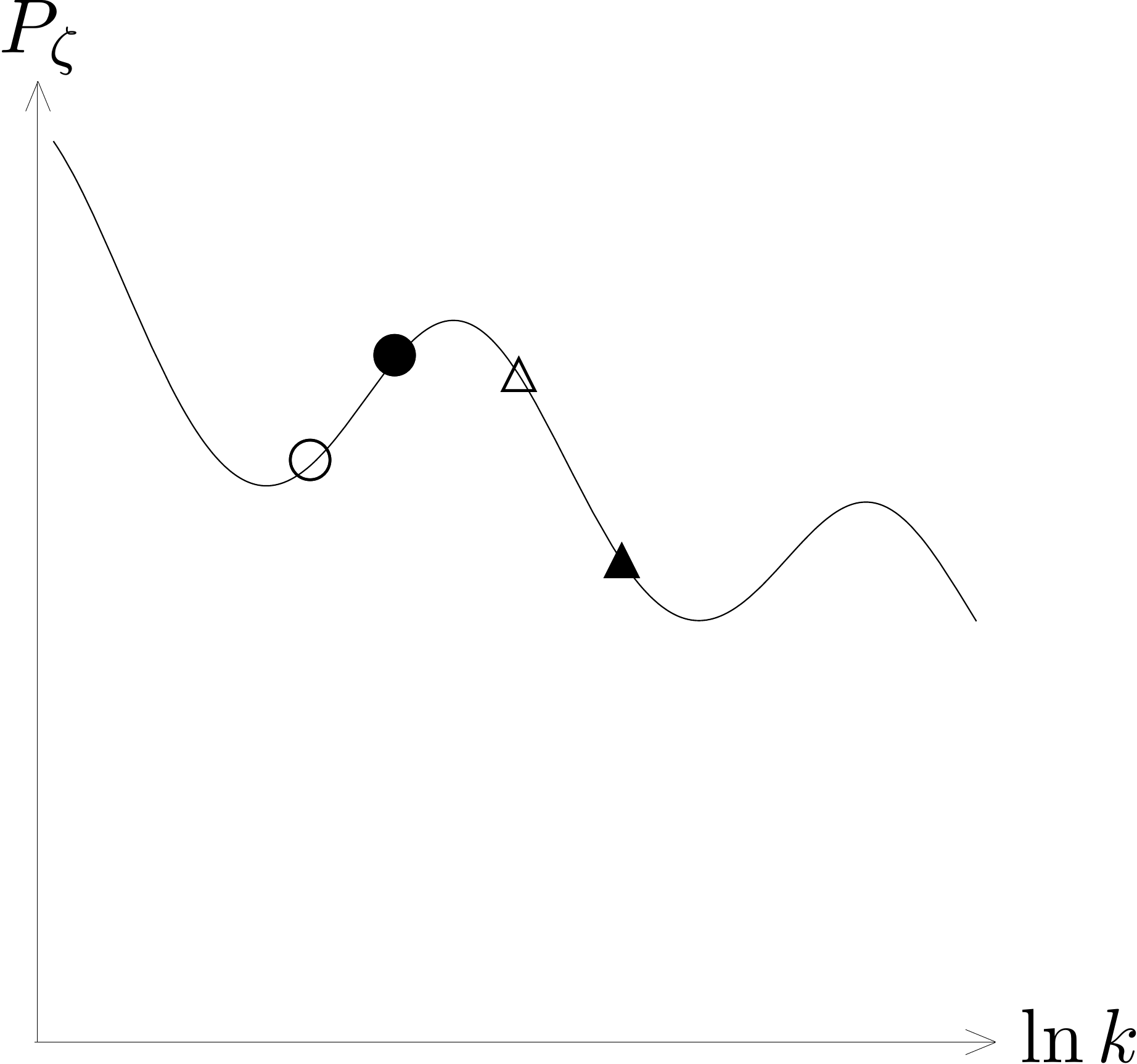}
 \caption{The curvature perturbation spectrum generated from the inflaton
 potential in Fig.~\ref{fig:im_pot}.}
 \label{fig:im_sp}
\end{figure}

The spectrum of the curvature perturbation
can be computed,
\begin{equation}
 P_{\zeta} \simeq \frac{H^4}{4 \pi^2 \dot{\phi}^2}, 
 \label{n1spec-1}
\end{equation}
which should be estimated at the time of horizon
exit~$k = a H$. Note that, since $\dot{\phi}$ is determined by $V'$, the main contribution
to the power spectrum arises from the smooth part $V_0'$,  and that the $V_{mod}'$
results in small oscillations of the perturbation
spectrum as illustrated in Fig.~\ref{fig:im_sp}.
The oscillatory feature of the perturbation
spectrum arises mainly from $\dot{\phi}$ (i.e. $V'_{mod}$), and the modulations of $H$ (i.e. $V_{mod}$) 
is subdominant in the following analysis. 
The spectrum is blue (red) for scales exiting
the horizon when the inflaton is rolling along the potential whose
curvature is positive (negative). To be precise, since the
dropping of the Hubble parameter reddens the perturbation spectrum, the
resulting spectrum becomes blue tilted when the deceleration of the
inflaton (i.e. the positive curvature of the potential) is large
enough.

\begin{figure}[t!]
 \includegraphics[width=0.7\linewidth]{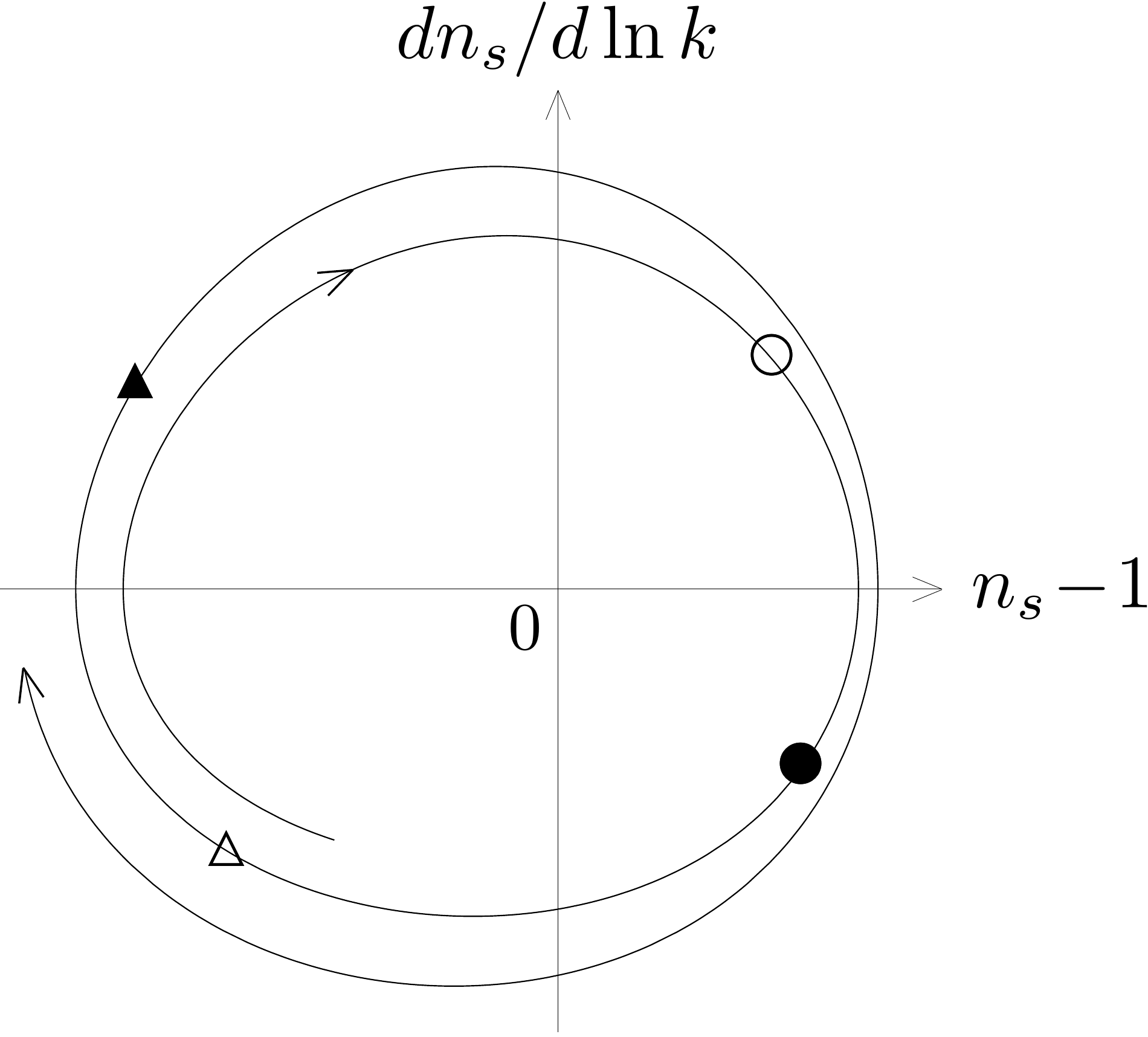}
 \caption{Spectral index and its running from the inflaton potential in
 Fig.~\ref{fig:im_pot}. }
 \label{fig:im_ti}
\end{figure}

Before computing the tilt of the spectrum, let us pause and comment on
the validity of the slow-roll approximations. 
Expressions for the spectral index and its running (and its running, and
so on) derived using the slow-roll approximations contain errors of
order of the approximate results multiplied by 
\begin{equation}
  \frac{1}{H} \frac{d}{dt} \simeq
  -  \frac{V'}{V} \frac{d}{ d\phi} .
 \label{slvalid}
\end{equation}
Therefore, as long as this is smaller than unity, results
obtained using the slow-roll approximations should be good estimates. 
Such condition is satisfied for the explicit cases we study later on,
since we require the observed scales to be generated within a half
oscillation period. 
For detailed analyses of inflation and primordial perturbations beyond
the slow-roll approximations, see
e.g. \cite{Stewart:1993bc,GarciaBellido:1996ke,Kinney:1997ne,Linde:2001ae,Kinney:2005vj,Tzirakis:2007bf,Kofman:2007tr,Kobayashi:2009nv}. 

Then the spectral index and its running can be estimated as follows,
\bea
 n_s - 1 & =  &\frac{d \ln P_{\zeta}}{d \ln k} \simeq
   - 6 \epsilon +2 \eta \non\\
  &\simeq& 
  - 6 \epsilon_0 + 2 \eta_0 
 - 6 \frac{V_0' V_{mod}'}{V_0^2} + 2 \frac{V_{mod}''}{V_0} ,
  \label{nssl} 
\eea
\bea
 \frac{d  n_s}{d \ln k} & \simeq &
 - 24 \epsilon^2 + 16 \epsilon \eta - 2 \xi \non\\ 
 \simeq  & &  \! \! \! \! \! \! \! \! \! 
 - 24 \epsilon_0^2 + 16 \epsilon_0 \eta_0 - 2 \xi_0 \non\\
 & & \! \! \! \! \! \! 
 - 24 \frac{V_0'^3 V_{mod}'}{V_0^4} + 8 \frac{V_0'^2 V_{mod}''}{V_0^3}
 - 2 \frac{V_0' V_{mod}'''}{V_0^2} ,
  \label{runsl}
\eea 
where
\begin{equation}
 \epsilon \equiv \frac{1}{2} \left(\frac{V'}{V}\right)^2 , \, \, 
 \eta \equiv \frac{V''}{V}, \, \, 
 \xi \equiv  \frac{V' V'''}{V^2}.
\end{equation}
In the final expressions, we have
kept contributions from the smooth 
potential~$V_0$ which are denoted by parameters with the subscript~$0$,
and also the leading oscillatory terms due to the
modulations~$V_{mod}$, by using the hierarchy~(\ref{hier})
and assuming that the inequality~(\ref{con2}) is well satisfied.
When the oscillatory terms are dominant, the tilt of the perturbation
spectrum repeatedly switches between red and blue,  
hence the spectral index~$n_s \! - \! 1$ as well as the running~$dn_s /
d\ln k$ change between positive and negative values.
This results in a clockwise motion in the $(n_s \! - \! 1)$ - $dn_s /
d\ln k$ plane as inflation proceeds, as illustrated in
Fig.~\ref{fig:im_ti}.
Given that the pivot scale exited the horizon when the inflaton was around
the filled circle in Fig.~\ref{fig:im_pot}, in one of the repeated
modulations, a blue tilted perturbation spectrum with negative running
is obtained. 

The hierarchy (\ref{hier}) implies that even if the modulations are so 
small that their effects on the Hubble parameter and the inflaton
velocity are subleading, they can still significantly contribute to the
spectral tilt of the perturbation spectrum. However, one should also
note that such hierarchy can lead to the breakdown of the conditions
(\ref{con1}) and (\ref{con2}) at other places along the inflaton
potential, thus considerably affecting the inflaton dynamics in the end. 
We will see in the following sections using explicit examples that
oscillatory modulations can satisfy the above conditions 
all through (or at least for a large fraction of) the inflationary period.

\section{Linear potential with periodic oscillations}
\label{sec:lin}
Now let us consider an explicit inflation model.
As a simple example, we consider curvature perturbations from
a linear inflaton potential~\cite{McAllister:2008hb,Takahashi:2010ky}
with superimposed periodic oscillations~\cite{Flauger:2009ab,Chen:2008wn} 
(cf.~Fig.~\ref{fig:im_pot}): 
\begin{equation}
 V(\phi) = \lambda \phi + \Lambda^4 \cos \left(\frac{\phi}{f}+ \theta \right),
 \label{linpot}
\end{equation}
where $\lambda$, $\Lambda$, $f$, and $\theta$ are constants.
The period of the oscillations is determined by
$f$, which we assume to be sub-Planckian, i.e. $f < 1$. This is
often the case in explicit microscopic constructions, and is also
compatible with our requirement that half of an oscillatory period roughly
corresponds to the time duration when the observed scales exited the
horizon. We also assume $V' > 0$, namely,
\begin{equation}
 \frac{\Lambda^4}{\lambda f} < 1, \label{1n1}
\end{equation}
which guarantees that the inflaton is not trapped by the modulations during inflation.
This condition is equivalent to (\ref{con2}).

Assuming that the slow-roll approximations (\ref{slow-roll})
are valid, the spectrum of the curvature perturbation
can be computed,
\begin{equation}
 P_{\zeta}
 \simeq  \frac{\lambda \phi^{3}}{12 \pi^2 } 
 \left(1 - \frac{\Lambda^4}{\lambda f} \sin \left(\frac{\phi}{f} + \theta \right)
 \right)^{-2} , 
 \label{n1spec}
\end{equation}
which should be estimated at the time of horizon exit. Note that  in
obtaining the right hand side we have omitted
modulations of $H$ (i.e. $V$) which is subdominant due to (\ref{1n1})
and $ f \ll \phi$ (since inflation occurs for super-Planckian~$\phi$). 

Let us now check whether we can use expressions for the spectral tilt
derived using the slow-roll approximations. 
(See also discussions below (\ref{slvalid}).) 
Upon its computations, one finds
\begin{equation}
 \left| -\frac{V'}{V} \frac{d}{d\phi} \right| \lesssim
  \frac{1}{\phi f}, \label{Mphif}
 \end{equation}
where the $f$ in the denominator of the right hand side comes from
differentiating the oscillatory part. 
As we will soon see, in our case $(\phi f)^{-1} $ is smaller than unity, 
at least until when the CMB scale exits the horizon, 
due to the requirement that the period of the oscillations be large
enough to encompass the observed scales. 
Therefore we trust results derived using the slow-roll
approximations. 

Then the spectral index and its running can be estimated as follows,
\begin{align}
 n_s - 1 
 &\simeq - \frac{1}{\phi^2} 
 \left( 3 + \frac{2 \Lambda^4 \phi}{\lambda f^2}
 \cos \left( \frac{\phi}{f} + \theta \right)\right), \label{n1ns}
\end{align}
\begin{align}
 \frac{d  n_s}{d \ln k} 
 &\simeq - \frac{1}{\phi^4} 
 \left( 6 + \frac{2 \Lambda^4 \phi^2}{\lambda f^3}
 \sin \left( \frac{\phi}{f} + \theta \right)\right). \label{n1run}
\end{align}
In the final expressions~(\ref{n1ns}) and (\ref{n1run}), we have only
kept the leading oscillatory terms. 
One clearly sees that the spectral index and its running oscillate with
growing oscillation amplitudes as inflation proceeds and $\phi$ becomes
smaller. (The relative amplitude compared to the non-oscillatory part becomes
smaller.)
Note that the power of $f (< 1)$ appearing in Eqs.~(\ref{n1ns}) and (\ref{n1run}) are larger than
that of Eq.~(\ref{n1spec}), and that is why the spectral index and its running can be
significantly modified without affecting the power spectrum too much. One can easily check
that the conditions (\ref{con1}) - (\ref{con4}) can be satisfied for a certain choice of $f$. 

Let us now set the parameters of the potential so that the spectral
index and its running at the pivot scale realizes the central values of
the bounds from WMAP 7-yr~\cite{Larson:2010gs}:
\begin{equation}
\begin{split}
 \label{WMAP7}
 n_s & = 1.076 \pm 0.065 ,  \\ 
 d n_s / d\ln k &= -0.048 \pm 0.029,
\end{split}
\end{equation}
(both at 68\% CL) when both the running and the tensor mode
perturbations are allowed to vary, in addition to the spectral
index. (The bound on the tensor-to-scalar ratio here is $r < 0.49$ (95\%
CL).)  Note that the predicted power spectrum is not exactly the one used 
in the analysis by the WMAP team. However, since the period of oscillations
is sufficiently long compared to the observed CMB scales, such a difference
will not significantly change the likelihood distribution. For simplicity we use
the above values in the following analysis. 

We set the pivot scale to have exited the horizon when the inflaton
field was at around $\phi_0 = 10$, which corresponds to about
50 e-folds before inflation ended.
For a linear potential (without modulations),
the number of e-folds obtained within the field range~$\Delta \phi$
starting from $\phi_0$ is $\Delta \mathcal{N} \simeq
\phi_0 \Delta \phi $. Here we set $f$ by choosing
the half period of oscillation from the pivot scale to correspond to
about $\ln l^{\mathrm{CMB}}_{max} \simeq 7.5$
e-foldings with $ l^{\mathrm{CMB}}_{max} \sim 2000 $, 
i.e. $\pi f = \Delta \mathcal{N} / 2  \phi_0 = 0.75 $. 
(Note that requiring the observed scales to lie within one oscillation
period, i.e. $\Delta \mathcal{N} \gtrsim 10$, leads to $( \phi_0 f)^{-1}
< 1$ which validates the analytic expressions derived using
slow-roll approximations, as was discussed below
(\ref{slvalid}) and (\ref{Mphif}).\footnote{The condition $( \phi f)^{-1} < 1$ is
satisfied until about 10 e-foldings before the end of inflation.})
Then the remaining parameters $\lambda$, $\Lambda$, $\theta$ are determined
so that the perturbation spectrum at the pivot scale realizes the COBE
normalization~\cite{Komatsu:2010fb} $P_\zeta = 2.43 \times 10^{-9}$, as
well as the central values of (\ref{WMAP7}). We choose $ \lambda \approx
1.99 \times 10^{-10}$, $\Lambda \approx 1.77 \times 10^{-3} $, $\theta
\approx -2.03$, and the inflaton field value when the pivot scale exits
the horizon as $\phi \approx 9.98$. We note that the modulations'
amplitude~$\Lambda$ is of order the GUT scale.

\begin{figure}[t!]
 \includegraphics[width=0.9\linewidth]{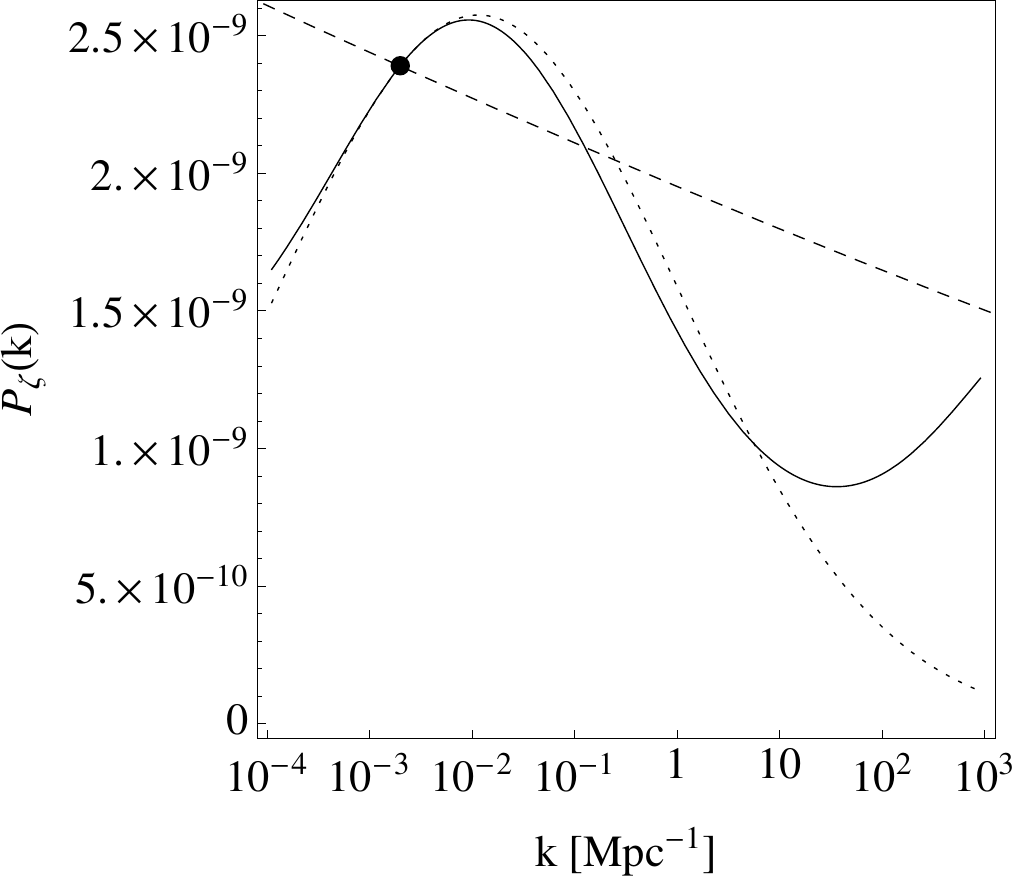}
 \caption{Curvature perturbation spectrum (black solid line) from the
 inflaton potential~(\ref{linpot}), realizing central values of
 (\ref{WMAP7}) at the pivot scale~$k_0 = 0.002 \mathrm{\, Mpc^{-1}}$
 which is denoted by the filled circle. Spectrum with constant
 running~(\ref{simpleex}) extrapolated from values at the pivot scale
 is shown as the black dotted line. Spectrum from a linear potential
 without any modulation is shown as the black dashed line for comparison.} 
 \label{fig:n1_spectrum}
\end{figure}
\begin{figure}[t!]
\includegraphics[width=0.9\linewidth]{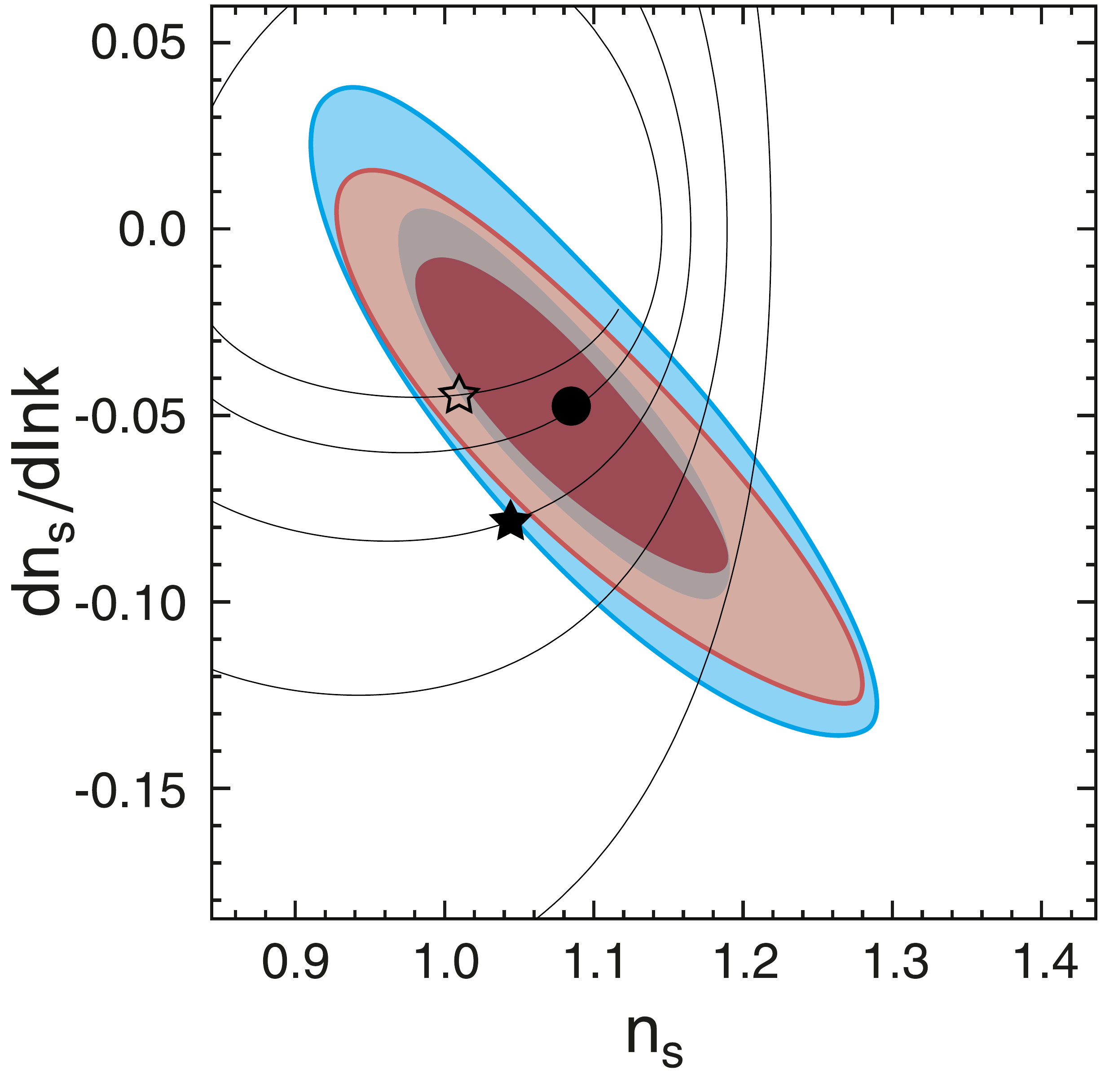}
 \caption{Variation of the spectral index~$n_s$ and its running~$dn_s /
 d\ln k$ as inflation proceeds along the modulated
 potential~(\ref{linpot}). The plot starts from 70 e-folds before the
 end of inflation (end point of the innermost orbit), up to $\sim$ 10
 e-folds before the end of inflation where most part of the orbit goes
 beyond the plotted region. Filled circle corresponds to the pivot scale
 (about $50$ e-folds before the end of inflation). The scale which
 exited the horizon 15 e-folds before (after) the pivot scale is
 indicated by an open (filled) star. Superimposed contours are
 constraints from 5-year WMAP (blue) and from 5-year WMAP+BAO+SN (red)
 (68\% and 95\% CL), when tensor moder perturbations, spectral tilt, and
 running index are allowed to vary~\cite{Komatsu:2008hk}.}
 \label{fig:n1_tilt}
\end{figure}

With these parameter sets, we have numerically solved the equation of
motion of the inflaton and the Friedmann equation, and computed the
curvature perturbation spectrum 
$P_{\zeta} = \left. \frac{H^4}{4 \pi^2
\dot{\phi}^2}\right|_{k=aH}$, 
which is shown in Fig.~\ref{fig:n1_spectrum}. 
Oscillatory features are clearly seen in the spectrum.
For comparison, we also plotted a spectrum with constant
running~$\alpha$, 
\begin{equation}
 P_{\zeta} (k) = P_{\zeta} (k_0) \left(\frac{k}{k_0}\right)^{n_s (k_0) -
  1 + \frac{1}{2}\alpha \ln \frac{k}{k_0} }, \label{simpleex}
\end{equation}
extrapolated from values at the pivot scale~$k_0 = 0.002 \mathrm{\,
Mpc^{-1}}$. The CMB observations probe scales from $k \sim 10^{-4}$
Mpc$^{-1}$ down to $\sim 0.1$ Mpc$^{-1}$, and within this range these
two lines are almost degenerate. The difference becomes evident only at
small scales~$k \gtrsim 10$~Mpc$^{-1}$, which may be probed by the
future proposal of the 21 cm
observations~\cite{Lewis:2007kz,Tegmark:2008au}. 
We also plot the varying spectral index 
and its running in Fig.~\ref{fig:n1_tilt}. For each 
oscillation in the inflaton potential, the trajectory in the phase space
passes through the WMAP constrained region. As inflation proceeds,
i.e. towards smaller scales, the radius of the orbit increases and also
its center shifts towards smaller~$n_s$.  


\section{Power-Law Potential with Periodic Oscillations}
\label{sec:power-law}

All the previous discussions can be straightforwardly extended to
large-field inflation with general power-law potential (see
e.g. \cite{Feng:2003mk,Wang:2002hf,Kaloper:2008fb} for models which
exhibit quadratic inflaton potentials with superimposed periodic oscillations),
\begin{equation}
 V(\phi) = \lambda \phi^n + \Lambda^4 \cos \left( \frac{\phi}{f} +
					      \theta \right).
 \label{powerlawpot}
\end{equation}
The condition equivalent to (\ref{1n1}), which guarantees $V' > 0$ is now
\begin{equation}
 \frac{\Lambda^4}{n \lambda \phi^{n-1} f} < 1, \label{condgenn}
\end{equation}
but one should note that for $n>1$, this may break down as $\phi $
becomes smaller and the inflaton can be trapped in a local minimum due
to the modulations.  

\begin{figure}[t!]
 \includegraphics[width=0.9\linewidth]{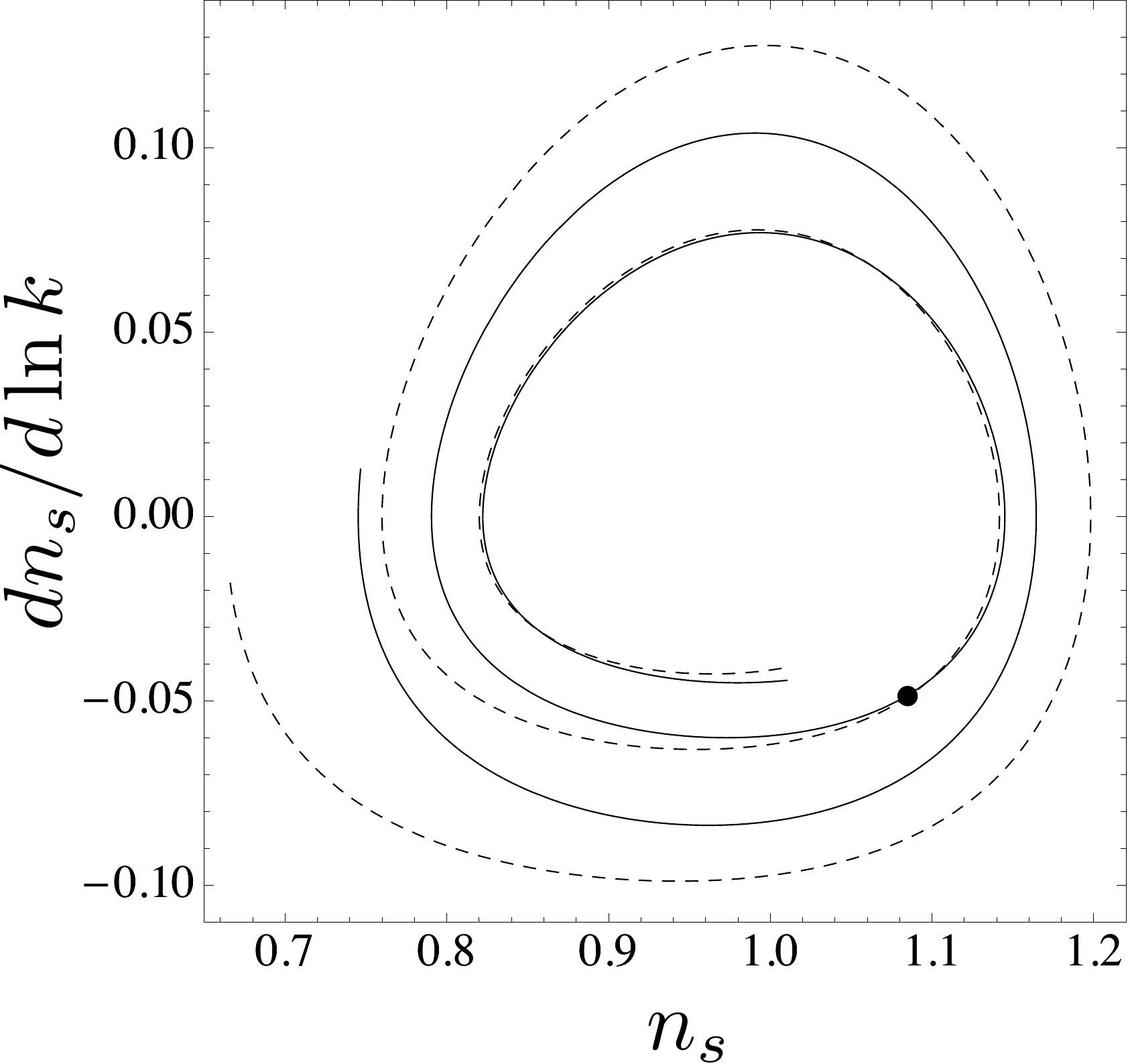}
 \caption{Variation of the spectral index~$n_s$ and its running~$dn_s /
 \ln k$ for linear (solid) and quadratic (dashed) potentials with
 modulations. Filled circle indicates when the pivot scale exited the
 horizon. Both trajectories are plotted from 15 e-folds before the pivot
 scale exited the horizon (end points of the innermost orbits),
 up to 20 e-folds after the pivot scale (end points of the outermost orbits).}
 \label{fig:n1n2_tilt}
\end{figure}

The approximate expressions for the perturbation spectrum and its
spectral tilt can be derived in a similar fashion,
\begin{equation}
 P_{\zeta} 
 \simeq  \frac{\lambda \phi^{n+2}}{12 \pi^2 n^2  } 
 \left(1 - \frac{\Lambda^4}{n \lambda f \phi^{n-1}} \sin
  \left(\frac{\phi}{f} + \theta \right)  \right)^{-2},
 \label{genspec}
\end{equation}
\begin{multline}
 n_s - 1  \\
 \simeq - \frac{1}{\phi^2} 
 \left( n(n+2) + \frac{2 \Lambda^4 }{\lambda f^2 \phi^{n-2}}
 \cos \left( \frac{\phi}{f} + \theta \right)\right),
 \label{genns}
\end{multline}
\begin{multline}
 \frac{d  n_s}{d \ln k} \\
 \simeq - 2 n \frac{1}{\phi^4} 
 \left( n (n+2) + \frac{\Lambda^4 }{\lambda f^3 \phi^{n-3} } 
 \sin \left( \frac{\phi}{f} + \theta \right)\right).
 \label{genrun}
\end{multline}
One can see that the spreading of the oscillation amplitudes of $n_s\!
-\! 1$ and $dn_s / d\ln k$ as inflation proceeds is more drastic for
larger~$n$, since oscillations with constant amplitude as in
(\ref{powerlawpot}) become relatively more significant at small $\phi$. 
For comparison, we have plotted oscillating trajectories in the 
$(n_s \! - \! 1)$ - $dn_s /d\ln k$ plane for the linear ($n=1$) and
quadratic ($n=2$) potentials with modulations in
Fig.~\ref{fig:n1n2_tilt}. The parameters of the potentials are chosen in
order to realize the central values of (\ref{WMAP7}) and the COBE
normalization at the pivot scale. The case with the linear
potential was discussed in the previous section. 
For the quadratic potential, the pivot scale is set to have exited the
horizon at around $\phi_0 = \sqrt{200} $, and $f$ is chosen to realize a half
oscillation period lasting for about 7.5 e-foldings from the pivot scale,
i.e. $\pi f =  \Delta \mathcal{N} / \phi_0  \approx 1.06 $. 
The remaining parameters are chosen as $\lambda \approx 1.95
\times 10^{-11} $, $\Lambda \approx 2.52 \times 10^{-3} $, $
\theta \approx -2.06$, and the inflaton field value when the pivot
scale exits the horizon as $\phi \approx 14.1$.\footnote{For this
parameter set, the breakdown of the condition~(\ref{condgenn}) happens
at around the end of large-field inflation, hence we disregard the
trapping of the inflaton in a local minimum due to the modulations.
However, modulations do flatten the potential towards small~$\phi$, and
extend the period of inflation: The number of e-folds
obtained from when $\phi_0 \approx \sqrt{200}$ until the end of
inflation is about 50 for a quadratic potential without modulations, but
in this case about 57. Such increase/decrease of e-foldings
due to modulations may have interesting consequences, see discussions in
Sec.~\ref{sec:conc}.}

Normalized at the pivot scale, one sees that the trajectories for the
quadratic potential, compared to that of the linear case, starts from
an orbit with smaller radius and grows to a larger orbit towards the end
of inflation. 

\section{Discussions and Conclusions}
\label{sec:conc}

Compared to cases with constant running of the spectral index, the
perturbation spectrum from periodic modulations which we have studied is
enhanced at small scales, thus does not delay structure
formation~\cite{Yoshida:2003wf}. 
Throughout this work we have assumed that the slow-roll approximations
are valid during inflation. 
If we consider an extreme case where the slow-roll approximations are
violated temporally during inflation, a perturbation spectrum with large
oscillations at the corresponding scales is realized.  
Such a spectrum may be used to account for the anomalous suppression at
the quadrupole (see \cite{Bennett:2010jb,Kawasaki:2003dd} and references
therein) as well as the missing satellite and/or core-cusp
problems~\cite{Kamionkowski:1999vp,Dalcanton:2000hn}.

We focused on a case that a sizable running of the spectral index is
generated due to the modulations. 
We may use such a structure to modify the prediction on the spectral
index to make an inflation model consistent with observations. One
example is chaotic inflation with a quartic coupling, which is strongly
disfavored by the WMAP and other observations because the
tensor-to-scalar ratio is too large for the predicted spectral
index. 
The tension can be relaxed if modulations terminate inflation earlier
compared to the case without modulations. Then the inflaton field value
when the CMB scale exited the horizon is corrected, modifying
predictions on the spectral index and the amplitude of the tensor mode
perturbations.  
Another possibility is that the modulations directly increase the
spectral index. However for this to happen, the amplitude of the
modulations should decrease towards smaller values of the inflaton,
otherwise the inflaton would get trapped by one of 
the modulations as inflation proceeds. 

In this paper, we have explored the possibility of periodic modulations
to the inflaton potential giving rise to large running of the spectral
index at all scales. 
Due to the oscillatory modulations, the spectral index~$n_s \!  - \! 1$
as well as its running~$dn_s / d \ln k$ switch between positive and
negative values. Such behavior results in distinct features in the
perturbation spectrum at small scales, which may be probed by future
21~cm observations. 
As explicit examples, we have studied inflation models with power-law
potentials with superimposed periodic oscillations.
We showed that the modulations can produce a running spectral index,
while having small impact on the overall inflaton dynamics. 
In order for the cosmological observables to lie at the central values
of the WMAP bounds, modulations with amplitudes of order the GUT scale
were considered. 
Especially in large-field models, as the inflaton travels
super-Planckian field ranges, modulations in the potential can show up
with scales determined by the microphysics governing inflation. 
Concrete constructions of such cases in the context of string theory were
performed in \cite{McAllister:2008hb,Flauger:2009ab}, which realized
linear inflaton potentials with superimposed oscillations. 

Though we focused on modulations with a large oscillation period
encompassing the entire CMB scale, cases with small periods can also have
interesting consequences, such as oscillations of the perturbation
spectrum within the observed
scales~\cite{Hamann:2008yx,Wang:2002hf,Pahud:2008ae}. Very small
oscillations whose period is sufficiently shorter than one e-fold during
inflation, can further lead to resonance effects in the power spectrum
as well as in the non-Gaussianity, as was studied 
in~\cite{Flauger:2009ab,Chen:2008wn,Hannestad:2009yx,Flauger:2010ja}. 

Large running of the spectral index, if confirmed, may suggest the
presence of abundant features in the inflaton potential, indicating
connections with the microphysics. 
It would be interesting to examine what we can say about the underlying
physics if future observations detect combinations of signals, such as
tensor modes and a running spectral index.

\begin{acknowledgments}
 We would like to thank Juerg Diemand, Raphael Flauger, Masahiro
 Kawasaki, Hironao Miyatake, Shinji Mukohyama, and Naoki Yoshida for
 helpful conversations. The work of T.K. was supported by Grant-in-Aid for
 JSPS Fellows No.~21$\cdot$8966. This work was supported by World
 Premier International Research Center Initiative (WPI Initiative),
 MEXT, Japan. 
\end{acknowledgments}

\appendix


\end{document}